\documentclass[runningheads]{llncs}
\usepackage{graphicx}
\usepackage{microtype}
\usepackage[T1]{fontenc}
\usepackage{lmodern}
\usepackage{breakcites}

\usepackage{xurl}
\newcommand{\repeatthanks}{\textsuperscript{\thefootnote}}
\begin{document}
\title{PreprintResolver: Improving Citation Quality by Resolving Published Versions of ArXiv Preprints using Literature Databases 
}

\titlerunning{PreprintResolver: Resolving Published Versions of ArXiv Preprints}
%
\author{Louise Bloch\inst{1,2,3}\thanks{these authors contributed equally to the work} \orcidID{0000-0001-7540-4980} \and Johannes Rückert\inst{1}\repeatthanks\orcidID{0000-0002-5038-5899} \and
Christoph M. Friedrich\inst{1,2}\orcidID{0000-0001-7906-0038}}
\authorrunning{L. Bloch et al.}
%
\institute{Department of Computer Science, University of Applied Sciences and Arts Dortmund, Emil-Figge-Str. 42, 44227 Dortmund, Germany \email{\{louise.bloch,johannes.rueckert,christoph.friedrich\}@fh-dortmund.de}\\ \and
Institute for Medical Informatics, Biometry and Epidemiology (IMIBE), University Hospital Essen, Hufelandstraße 55, 45122 Essen, Germany \and Institute for Artificial Intelligence in Medicine (IKIM), University Hospital Essen, Hufelandstraße 55, 45122 Essen, Germany}
\maketitle              
\begin{abstract}
The growing impact of preprint servers enables the rapid sharing of time-sensitive research. Likewise, it is becoming increasingly difficult to distinguish high-quality, peer-reviewed research from preprints. Although preprints are often later published in peer-reviewed journals, this information is often missing from preprint servers. To overcome this problem, the PreprintResolver was developed, which uses four literature databases (DBLP, SemanticScholar, OpenAlex, and CrossRef / CrossCite) to identify preprint-publication pairs for the arXiv preprint server. The target audience focuses on, but is not limited to inexperienced researchers and students, especially from the field of computer science. The tool is based on a fuzzy matching of author surnames, titles, and DOIs. Experiments were performed on a sample of 1,000 arXiv-preprints from the research field of computer science and without any publication information. With 77.94~\%, computer science is highly affected by missing publication information in arXiv. The results show that the PreprintResolver was able to resolve 603 out of 1,000 (60.3~\%) arXiv-preprints from the research field of computer science and without any publication information. All four literature databases contributed to the final result. In a manual validation, a random sample of 100 resolved preprints was checked. For all preprints, at least one result is plausible. For nine preprints, more than one result was identified, three of which are partially invalid. In conclusion the PreprintResolver is suitable for individual, manually reviewed requests, but less suitable for bulk requests. The PreprintResolver tool\footnote{PreprintResolver tool: \url{https://preprintresolver.eu}, Available from 2023-08-01} and source code\footnote{PreprintResolver code: \url{https://gitlab.com/ippolis_wp3/preprint-resolver}, Accessed: 2023-07-19} is available online.
\keywords{Preprint  \and arXiv \and Publication \and Research Quality \and Digital Library.}
\end{abstract}

\section{Introduction}
Preprints are scientific manuscripts that have been uploaded by their authors to public servers (so-called preprint-servers) without being subjected to any peer review process \cite{Berg2016DefinitionPreprint}. The original idea of preprints was the publication of research before or during submission to a peer-reviewed journal or conference. More recently, however, most preprint servers also include research that has been submitted after it has been accepted or published in peer-reviewed venues.

ArXiv \cite{Ginsparg1994ArxivOriginalPublication,Ginsparg2011Arxiv20years} was one of the first internet-based preprint servers developed in 1991 to share preprints in the field of physics. Since then, arXiv has expanded to include the fields of mathematics, computer science, quantitative biology, quantitative finance, statistics, electrical engineering and systems science, and economics. Recently, there have been multiple preprint servers, most of them focusing on specific research areas, e.g., bioRxiv \cite{Sever2019Biorxiv} in the field of biology, medRxiv \cite{Rawlinsonl2019MedRxiv} in the field of health sciences, the Social Science Research Network (SSRN) \cite{SSNR} in the field of social sciences and humanities, which later expands to science and engineering, or Humanities Commons \cite{HumanitiesCommons} in the field of humanities.

In comparison to traditional publication, the release of preprints promises a faster publication of time-sensitive research \cite{Abdill2019HighPublicationRateOfPreprints}. In addition, the benefits of preprints include higher attraction \cite{Kelly2018IncreasedAttractionByPreprints} and potentially more citations \cite{Davis2007ArxivLeadsToHigherCitationCounts,Feldman2018citationOfArxivPreprintMatchingOfHighInfluentialCSConferences,Fraser2020BiorxivPreprintCitations,Fu2019MetaResearchPreprintsAssociatedWithMoreCitation,Serghiou2018CitationsOfStudiesPostedAsPreprintd}, faster feedback from the research community as well as open access publication. While the benefits initially outweigh the disadvantages for the submitter, the disadvantages primarily affect the research community. Due to the lack of peer review, the quality of preprints is not guaranteed which increases the risk of fraudulent preprints and preprints with low research quality. This makes it more difficult to identify quality research, especially for inexperienced researchers. In particular, there is a risk of citing outdated versions of research possibly containing altered or incorrect information \cite{Fraser2020BiorxivPreprintCitations}. This risk is increased because although preprints are often submitted, accepted, and published in peer-reviewed journals or conference proceedings \cite{Abdill2019HighPublicationRateOfPreprints}, preprint servers often lack a link to those updated publications 
\cite{Cabanac2021PreprintPublicationLinker,Eckmann2023PreprintMatch,Lin2020ResolvePublishedArxivPreprintsCS}. 

In this work, the PreprintResolver was developed to resolve published versions of arXiv preprints using four databases. It can help researchers to find the latest published version of preprints. The tool focuses on, but is not limited to, a target group of inexperienced researchers and students in the field of computer science. 

\section{Related Work}
There are a few tools with similar ideas. For example, the SAGE Rejected Article Tracker\footnote{SAGE Rejected Article Tracker: \url{https://github.com/sagepublishing/rejected_article_tracker_pkg}, Accessed: 2023-07-19} presented in \cite{Hails2021SAGERejectedArticleTracker} was developed to track papers rejected by journals, but can also be used to resolve preprints. The Python library requires the input of titles and authors and identifies published versions using CrossRef \cite{Rachael2014CrossRef}. Levenshtein distance \cite{Levenshtein1966LevenshteinDistance} and a logistic regression model were used to determine whether publications were fuzzy matches. The tool is not specialised in resolving preprints which increases the complexity of identifying quality research for inexperienced researchers.

In \cite{Eckmann2023PreprintMatch} the tool PreprintMatch\footnote{PreprintMatch: \url{https://github.com/PeterEckmann1/preprint-match}, Accessed: 2023-07-19} is presented that matches preprints uploaded on the BioRxiv and MedRxiv preprint servers to PubMed \cite{canese2013pubmed} publications. The tool is based on database dumps. It matches the titles and abstracts using a word vector representation to handle semantic changes. The vectors are ranked by cosine similarity and the top 100 results are compared using the Jaccard similarity \cite{Jaccard1912JaccardIndex} of author names and a Support Vector Machine (SVM) \cite{Cortes1995SVM} trained on a hand-selected dataset. The main differences are the focus on the biomedical domain and the use of database dumps with the risk of outdated information.

A tool named PreprintPublicationLinker\footnote{PreprintPublicationLinker: \url{https://github.com/gcabanac/preprint-publication-linker}, Accessed: 2023-07-19} uses more recent data by requesting the CrossRef Application Programming Interfaces (API) \cite{Cabanac2021PreprintPublicationLinker}. It matches preprints from medRxiv with the CrossRef API. The implementation includes a fuzzy comparison of titles, authors, ORCIDs, and publication time. It was tested on a corpus of preprints related to COVID-19 and reached an accuracy of 91.5~\% (sensitivity: 90.9~\%, specificity: of 91.9~\%). The main difference to this work is the focus on the medical domain and the use of a single literature database. 

For the arXiv preprint server, which is the focus of this work, the Bibliographic Explorer\footnote{arXiv Bibliographic Explorer: \url{https://github.com/mattbierbaum/arXiv-bib-overlay}, Accessed: 2023-07-19} exists. It provides information about published versions and citations directly as an overlay of the arXiv website. It is based on the databases SemanticScholar \cite{Ammar2018SemanticScholar,Kinney2023SemanticScholar,Lo2020SemanticScholar}, Google Scholar \cite{GoogleScholar}, CrossRef / CrossCite, NASA Astrophysics Data System (ADS) \cite{accomazzi2015ADS}, and the Inspire HEP API \cite{Moskovic2021InspireHEPAPI}. The tool includes no fuzzy matching of database and preprint information but provides links to the database results directly associated with the \texttt{arXiv-id} or DOI. This leads to links to the original arXiv preprints, even though a published version exists, which can be confusing for inexperienced researchers. 

Additionally, some case studies use matching algorithms to identify preprint-publication pairs. For example, \cite{Lin2020ResolvePublishedArxivPreprintsCS} presents a study investigating how many, and which arXiv preprints were published in the field of computer science. The matching was based on crawled data from CrossRef \cite{Rachael2014CrossRef} and Digital Bibliography and Library Project (DBLP) \cite{Ley2002DBLP}. For papers with unchanged titles, a fuzzy matching of first authors and titles was implemented. A Bidirectional Encoder Representations from Transformers (BERT) \cite{Devlin2019BERT} model was trained to match preprints and publications with changed titles. The model was trained using a dataset containing arXiv version data as well as data from a CrossRef search.

In \cite{Lariviere2014MatchingOfArxivEprintsAndJournal}, arXiv preprints are matched to publications in the Web Of Science (WoS) library using a fuzzy matching of titles, journals mentioned in arXiv comments, and first authors. Afterwards, the first characters of the abstracts are compared. During the analysis, 63.7~\% of the preprints were resolved, but the rate varies by discipline. For example, in computer science, less than 20~\% were resolved. One reason may be the impact of conferences in this field \cite{Lisee2008ComputerScienceConferencesMoreImportent}.

For an analysis of citations and altmetrics in preprints, a matching between bioRxiv preprints and CrossRef and Scopus \cite{Scopus} was implemented in \cite{Fraser2020BiorxivPreprintCitations}. It uses the bioRxiv property of CrossRef, a scan of the bioRxiv websites, and a fuzzy matching including the authors, titles, and abstracts using Scopus as a database. The algorithm resolved 67.6~\% of the preprints. 

In \cite{sutton2017popularityOfArxivInCSMapping} a backward resolving of arXiv preprints is implemented for highly influential computer science conferences. All publications of 63 conferences are identified using the DBLP. An arXiv dump was used to identify corresponding preprints using an exact matching of titles and one author. For 56~\% of conference articles, a preprint was found.

To the best of our knowledge, there is currently no tool that helps inexperienced researchers identify the most recent publication of an arXiv preprint. The presented tool uses multiple literature databases to resolve preprints from different disciplines and returns the BibTeX citations that can be imported directly into bibliographies. As the tool communicates with the APIs of four literature databases, the data is up to date and no time-intensive downloading of database files is required. The tool\footnote{PreprintResolver tool: \url{https://preprintresolver.eu}, Available from 2023-08-01} and the source code\footnote{PreprintResolver source code: \url{https://gitlab.com/ippolis_wp3/preprint-resolver}, Accessed: 2023-07-19} is available online.
\section{Literature databases}
This section introduces the databases used to obtain information about arXiv preprints and to find matching peer-reviewed publications. It has to be noted that there is a risk of incorrect or outdated information in all of the databases.
\subsection{arXiv}
Developed in 1991, arXiv \cite{Ginsparg1994ArxivOriginalPublication,Ginsparg2011Arxiv20years} was one of the first internet-based preprint servers. The initial aim was to share preprints in the field of physics. Later, arXiv expanded to the fields of mathematics, computer science, quantitative biology, quantitative finance, statistics, electrical engineering and systems science, and economics. Currently, arXiv contains more than 2.2 million preprints and has 2.6 billion total downloads \cite{Arxiv2023ArxivReport}. For linking between preprints and published research, authors are requested to add DOIs and additional information to the preprints after publication\footnote{arXiv add journal reference: \url{https://info.arXiv.org/help/jref.html}, Accessed: 2023-07-19}. Metadata from arXiv can be requested via a public API\footnote{arXiv API: \url{https://info.arXiv.org/help/api/user-manual.html}, Accessed: 2023-07-19}. 
\subsection{DBLP}
The DBLP computer science bibliography \cite{Ley2002DBLP} is a bibliographic library of scholarly literature in the computer science domain. In 2023, DBLP contains metadata of more than 6.7 million publications\footnote{DBLP record statistics: \url{https://dblp.org/statistics/recordsindblp.html}, Accessed: 2023-07-19}. 
As conferences and workshops have a high influence in the computer science domain \cite{Lisee2008ComputerScienceConferencesMoreImportent}, at the time of submission, 48.37~\% of the publications include conference and workshop papers, whereas 39.38~\% include journal articles\footnote{DBLP publication type statistics: \url{https://dblp.org/statistics/distributionofpublicationtype.html}, Accessed: 2023-07-19}. DBLP provides a public API where users can request publications, venues, and authors\footnote{DBLP API: \url{https://dblp.org/faq/How+to+use+the+dblp+search+API.html}, Accessed: 2023-07-19}. Among other information, requested metadata contains the title, authors, DOI, venue, year, and publication type.
\subsection{CrossRef / CrossCite}
CrossRef \cite{Rachael2014CrossRef} is a DOI registration agency launched in 2000. The original idea was to link research articles from different publishers to improve the citation resolution. CrossRef assigns and links unique identifiers to authors, works, research institutions, and funding. This makes it easier for researchers to find and cite quality research. Recently, CrossRef contains more than 147 million records\footnote{CrossRef: \url{https://www.crossref.org/06members/53status.html}, Accessed: 2023-07-19}, which were published in more than 120,000 journals and 102,000 conference proceedings. A publicly available REST-API of CrossRef is available online\footnote{CrossRef API: \url{https://www.crossref.org/documentation/retrieve-metadata/rest-api/}, Accessed: 2023-07-19} providing metadata and links between research objects.

\subsection{SemanticScholar}
SemanticScholar \cite{Ammar2018SemanticScholar,Kinney2023SemanticScholar,Lo2020SemanticScholar} is a literature database released in 2015 by the Allen Institute for Artificial Intelligence. It is based on a literature graph that links papers, authors and entities, and aims to help scientists discover and understand scientific literature. Publication metadata includes authors, titles, citation counts, venues, and publication years, among other information. The database contains more than 212 million research items\footnote{SemanticScholar: \url{https://www.semanticscholar.org/}, Accessed: 2023-07-19}. The publicly available API\footnote{SemanticScholar API: \url{https://www.semanticscholar.org/product/api}, Access: 2023-07-19} supports a limited number of queries per time. For this project, a key was requested which increases the number of requests. ArXiv-ids are often directly linked to published versions, and it is possible to request data directly using this id.
\subsection{OpenAlex}
OpenAlex \cite{Priem2022openalex} is a publicly accessible index of information on academic publications, authors, venues, institutions, and concepts which was launched in 2022. The research items in OpenAlex are linked using a graph database which contains more than 209 million academic publications \cite{Priem2022openalex} and is growing by about 50,000 per day \cite{Priem2022openalex}. Data can be requested via a publicly available REST-API\footnote{OpenAlex API: \url{https://docs.openalex.org/}, Accessed: 2023-07-19}. Among others, the metadata available for publications in OpenAlex contains authors, titles, doi, additional sources (such as arXiv), and citation counts. Currently it is not possible to request arXiv-ids directly in OpenAlex. 
\section{Methods}
\begin{figure}[ht!]
\includegraphics[width=\textwidth]{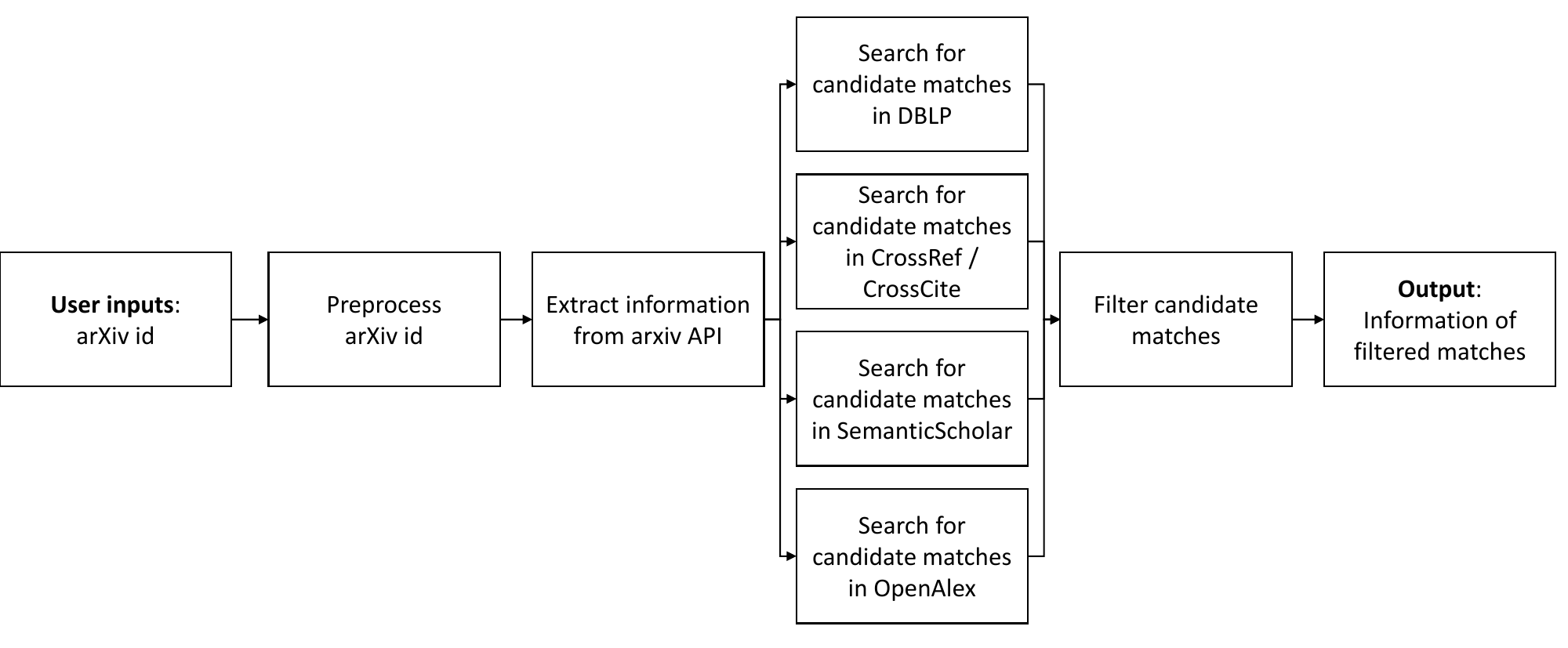}
\caption{Workflow used to resolve arXiv preprints based on four literature databases.} \label{fig:workflow}
\end{figure}

This section describes the workflow used to resolve arXiv preprints which is visualized in Fig. \ref{fig:workflow}.  

\subsection{User Input}
The tool is designed for, but not limited to, a target group of inexperienced researchers. The idea is, to make users aware of the differences between preprints and peer-reviewed research, leading to the requirement for good usability. For this reason, the user has to enter minimal input including only the \texttt{arXiv-id} into the interface.

\subsection{Preprocess Input}
In order to avoid errors, the first step of the pipeline is a preprocessing step. This includes removing leading and closing spaces. Second, as the input may contain the prefixes \texttt{"arxiv:"}, \texttt{"abs/"}, or \texttt{"pdf/"}, the last occurrence of these substrings is identified and all characters before and including these prefixes are removed. Additionally, \texttt{".pdf"} is deleted from the end of the input. Thus the resolver can also handle URLs in the form \url{https://arXiv.org/abs/{arXiv-id}} and \url{https://arXiv.org/pdf/{arXiv-id}.pdf}. A lower-case version of the extracted \texttt{arXiv-id} is used for further processing. 

\subsection{Extract Information from arXiv API}
Using the preprocessed \texttt{arXiv-id}, information from the publicly available arXiv-API is requested by \url{https://export.arXiv.org/api/query?id_list={arXiv-id}}. The response information is used to extract the latest arXiv version, authors, title, DOI, publishing date, categories, and additional information.
\subsection{Identify Candidate Matches in Literature Databases}
The \texttt{arXiv-id}, \texttt{title}, and \texttt{DOI} extracted from the arXiv API are used to identify candidate matches in the previously introduced literature databases.

\subsubsection{DBLP}
The DBLP computer science bibliography is requested for the \texttt{title} using the query: \url{https://dblp.org/search/publ/api?q={title}&format=json&h=5}. This query identifies the top five DBLP candidate matches. 

\subsubsection{CrossRef / CrossCite}
A two-level query is used to identify candidate matches using the CrossRef and CrossCite APIs\footnote{CrossCite API: \url{https://citation.crosscite.org/docs.html}, Accessed: 2023-07-19}.
\begin{enumerate}
    \item If available, the CrossCite API is first requested for the \texttt{DOI} using the request: \url{https://doi.org/{DOI}}.
    \item If no valid results are found in the first step, the top 10 CrossRef results are requested using the \texttt{title}: \url{https://api.crossref.org/works?query.bibliographic={title}&sort=score&rows=10}.
\end{enumerate}

\subsubsection{SemanticScholar}\sloppy
Information from SemanticScholar is requested in a three-level query. For all queries, the requested fields are: \texttt{title, authors, journal, venue, year, abstract, publicationTypes, externalIds, isOpenAccess, publicationDate, fieldsOfStudy, s2FieldsOfStudy, referenceCount, citationCount, influentialCitationCount}.
\begin{enumerate}
    \item The API is queried directly for the given \texttt{arXiv-id} using \url{https://api.semanticscholar.org/graph/v1/paper/ARXIV:{arXiv-id}}.
    \item  If no valid reference is found in Step 1, and if a DOI is available, the SemanticScholar API is requested for the \texttt{DOI} using \url{https://api.semanticscholar.org/graph/v1/paper/DOI:{DOI}}.
    \item If the previous steps find no valid matches, the API is requested for the \texttt{title}: \url{https://api.semanticscholar.org/graph/v1/paper/search?query={title}}.
\end{enumerate}
\subsubsection{OpenAlex}
The candidates from OpenAlex are requested in a three-level query.
\begin{enumerate}
    \item If available, the API is queried for the \texttt{DOI}: \url{https://api.openalex.org/works/doi={DOI}}.
    \item If the previous step shows no results, a search for the \texttt{title} is performed using \url{https://api.openalex.org/works?search={title}}. From this query, the results with a link to the original \texttt{arXiv-id} are first filtered.
    \item If no matches are identified in the previous steps, the initial candidate matches from Step 2 are filtered.
\end{enumerate}
\subsection{Filter Matches}
Two approaches have been implemented to filter the candidate matches. The first one is a weak filtering strategy, which is used when the results are identified in the database directly by the \texttt{arXiv-id}. The second strategy is a strong filtering which is used for the remaining cases. The weak filtering strategy is a subset of the strong filtering procedure.
\subsubsection{Weak Filtering}\label{Sec:WeakFiltering}
The idea behind weak filtering is that titles and authors can change during publication, so publications that are directly linked to the preprint in literature databases (SemanticScholar and OpenAlex Step 1) contain more evidence than fuzzy matching. The weak filtering contains the following steps:
\begin{enumerate}
    \item Exclusion of candidate matches where the preprint is identified as the publication.
    \item Exclusion of candidate matches without publication types, or venues.
    \item If the DOI is available in both papers, candidate matches with mismatching DOIs are discarded.
\end{enumerate}
\subsubsection{Strong Filtering}\label{Sec:StrongFiltering}
The matches which are identified without a direct link to the \texttt{arXiv-id} are filtered using the following criteria:
\begin{enumerate}
    \item The weak filtering strategy is applied.
    \item For the remaining candidate matches, the titles are compared using fuzzy matching. Titles are accepted if the ratio of the Levenshtein distance \cite{Levenshtein1966LevenshteinDistance} and the maximum number of characters in both titles is less than $0.05$.
    \item The last step is a comparison of the authors. This includes the removal of diacritical marks. Candidates are accepted if the ratio of successfully matched authors and the maximum number of authors of both papers exceeds $0.70$.
\end{enumerate}

\subsection{Output}
The PreprintResolver outputs structured information about the preprint and the candidate publications, as well as citation information. The citation information are provided in the BibTeX format and can thus be added directly to literature databases. Because authors sometimes publish papers with identical titles as journal and conference articles, the PreprintResolver can return multiple results for one database and the user has to choose which publication is most relevant.

\section{Experiments and Results}
In this section, some experiments are described which were performed to validate the functionality of the tool. The experiments are based on the arXiv Dataset\footnote{arXiv dataset: \url{https://www.kaggle.com/datasets/Cornell-University/arXiv}, Accessed: 2023-07-19} \cite{Arxiv2023ArxivDataset} that was published by Cornell University. The dataset was downloaded on 2023-01-19 and includes the metadata of $2,190,411$ preprints. Tab. \ref{Table:OverviewArxivDataset} summarizes the number of preprints, the number of preprints without any information about their publication, and the mean number of versions per preprint and research field. Publication information is the DOI and the journal reference. ArXiv allows authors to select multiple research fields for a preprint. To avoid duplicates in the evaluation process, this work focuses on the primary research field.

\begin{table}[t]
\caption{Overview of the arXiv dataset by primary research fields. Publication information is the DOI and journal reference. The dataset was downloaded on 2023-01-19.}\label{Table:OverviewArxivDataset}
\begin{center}

\begin{tabular}{|l|r|r|r|}
\hline
Research field extracted& \# preprints (ratio)&\# preprints W/O&Average\\
from primary arXiv category &&without&preprint\\
&&information (ratio)&versions\\
\hline\hline
Physics&1,234,491~~(56.36~\%)&287,269~(23.27~\%)&1.53\\
Mathematics&446,833~~(20.40~\%)&324,414~(72.60~\%)&1.72\\%
Computer Science&393,434~~(17.96~\%)&306,634~(77.94~\%)&1.55\\%
Quantitative Biology&24,887~~~(1.14~\%)&14,131~(56.78~\%)&1.49\\%
Quantitative Finance&9,785~~~(0.45~\%)&6,935~(70,87~\%)&1.69\\%
Statistics&41,161~~~(1.88~\%)&32,222~(78.28~\%)&1.75\\%
Electr. Eng. \& Systems Service&34,830~~~(1.59~\%)&26,712~(76.69~\%)&1.46\\%
Economics&4,990~~~(0.23~\%)&4,229~(84.75~\%)&1.76\\\hline
Overall&2,190,411~(100.00~\%)&1,002,546 (45.77~\%)&1.58\\
\hline
\end{tabular}
    
\end{center}
\end{table}

The dataset summary shows that the number of preprints differed between research fields. Most preprints are submitted in the initial area of arXiv -- physics ($1,234,491; 56.36~\%$). In computer science, the third most preprints ($393,434; 17.96~\%$) were uploaded. The ratios of preprints without any publication information also vary between disciplines. The ratio for the overall data set is $45.77~\%$. Only physics undercuts this ratio ($23.27~\%$). The highest ratio of $84.75~\%$ was reached in economics. Computer science reached the third highest ratio of $77.94~\%$. The mean number of versions per preprint differs between $1.46$ in electrical engineering and systems science and $1.76$ in economics. No clear association between the ratio of preprints without publication information and the mean number of versions was investigated.

As this tool focuses on the target group of inexperienced researchers in computer science, the experiments also do. The publication process for scientific journals and conference proceedings can take from a few months to years. To avoid biases, the experiments are based on all preprints with a first submission before January 2022. The computer science data set matching this criterion contains $327,320$ ($83.20~\%$ of preprints in computer science) preprints. The number of samples without any publication information is $248,671$ ($75.97~\%$) in this dataset. For evaluation, a random sample of $1,000$ preprints was drawn from this subset.

\begin{figure}[ht!]
\begin{center}
\includegraphics[width=0.6\textwidth]{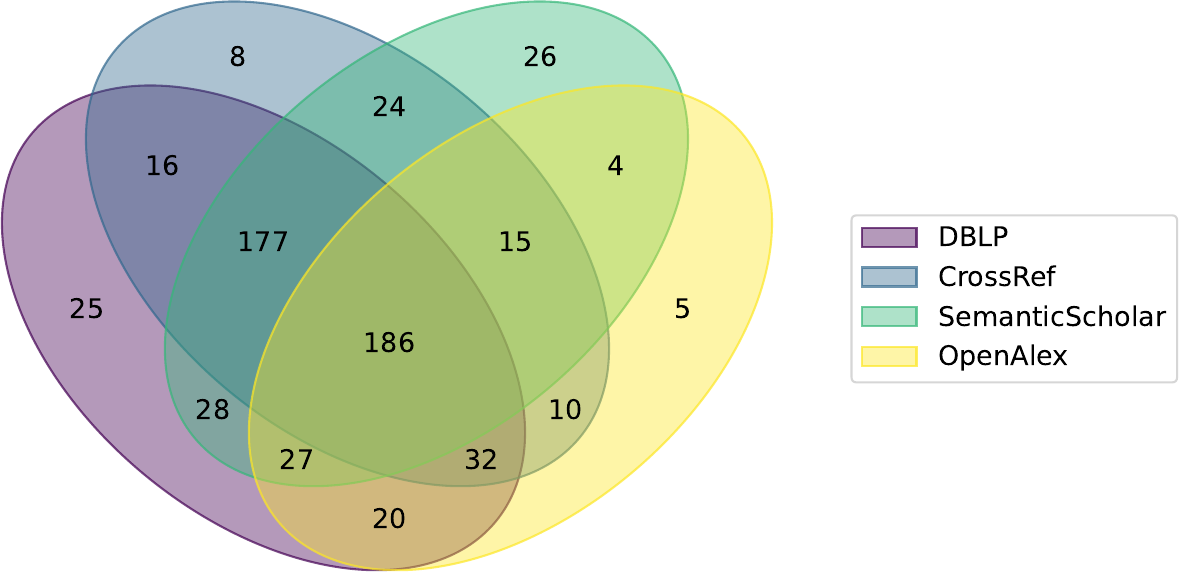}
\end{center}
\caption{Venn diagram of the literature database contributions in the PreprintResolver during resolving of a random sample of 1,000 preprints without any publication information in computer science.} \label{fig:venn}
\end{figure}

The results of the PreprintResolver are summarized in Tab. \ref{Table:SummaryExperimentalResults} and show that $603$ ($60.3~\%$) preprints were successfully resolved. Most preprints were resolved using DBLP ($51.1~\%$) followed by SemanticScholar ($48.7~\%$), CrossRef / CrossCite ($46.8~\%$), and OpenAlex ($29.9~\%$). The Venn diagram in Fig. \ref{fig:venn} shows, the contribution of the databases. $186$ ($18.6~\%$) of the preprints were found in all four databases. For all databases, there were preprints that are individually identified (DBLP: $25; 2.5~\%$, CrossRef / CrossCite: $8; 0.8~\%$, SemanticScholar: $26; 2.6~\%$, OpenAlex: $5; 0.5~\%$). Thus all databases contributed to the final result.

\begin{table}[t]
\caption{PreprintResolver results achieved by resolving a random sample of $1,000$ preprints without any publication information in computer science.}\label{Table:SummaryExperimentalResults}
\begin{center}
\begin{tabular}{|l|r|r|}
\hline
Database&\# preprints found&Resolving ratio\\
\hline\hline
DBLP&511&51.1~\%\\
CrossRef / CrossCite&468&46.8~\%\\
SemanticScholar&487&48.7~\%\\
OpenAlex&299&29.9~\%\\
\hline
Overall&603&60.3~\%\\
\hline
\end{tabular}
\end{center}
\end{table}

\begin{figure}[ht!]
\begin{center}
\includegraphics[width=0.6\textwidth]{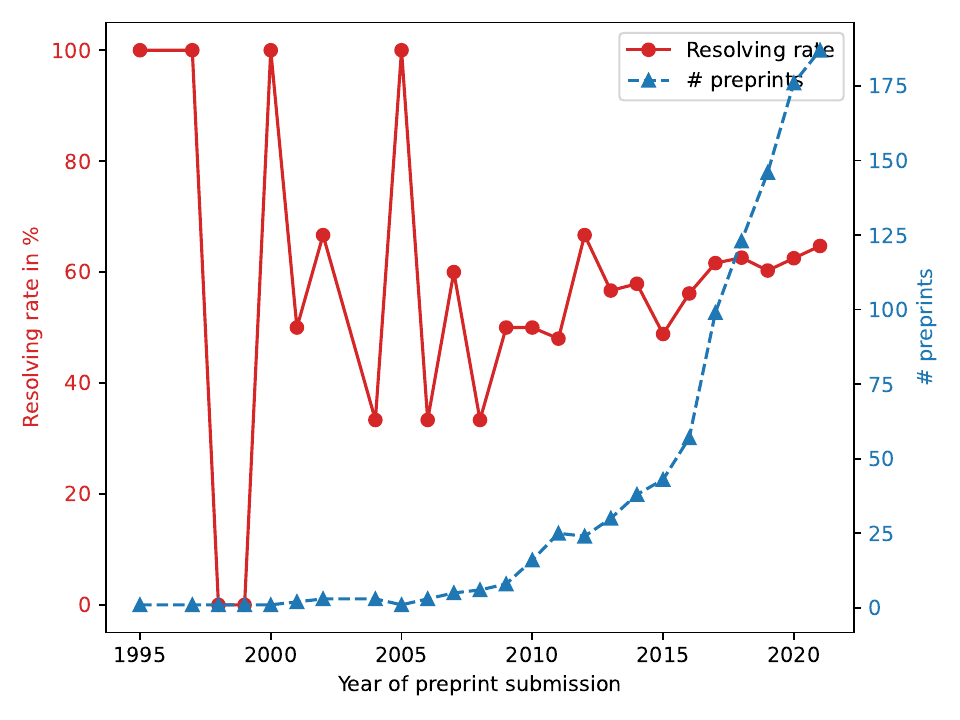}
\end{center}
\caption{Plot that shows the number of preprints per year and the resolving rate of the PreprintResolver in the investigated sample of 1,000 preprints in computer science without publication information.} \label{fig:resolving_rate_per_year}
\end{figure}

The plot in Fig. \ref{fig:resolving_rate_per_year} shows the number of preprints and the resolving rate over the years. Consistent with the increasing number of preprints the number of preprints in the sample is also increasing. The resolving rate between 1995 and 2009 shows strong fluctuations between $0~\%$ and $100~\%$. This can be attributed to the small number of samples (1995: 1; 2009: 8). From 2010, the samples increase and the resolving rate settles between $48.00~\%$ (2011) and $64.71~\%$ (2021). An increase in the resolving rate can be detected for recent preprints.

For validation, a random sample of 100 of the 603 results resolved by the PreprintResolver were manually checked by comparing the authors, titles, affiliations, and the beginning and end of the abstracts. The results show that the PreprintResolver identified at least one plausible match for all 100 preprints examined. For nine preprints, two different results were identified. Three of these results provide at least one incorrect or untraceable match. For four results it was not possible to determine which was the best match. For two results with two matches, the arXiv comments give a hint to the correct match. Of the 100 preprints examined, an unstructured arXiv comment provides information about the correct journal or conference proceedings for 46 preprints. For one preprint, a DOI has been added between dataset download and the resolving.

\section{Discussion and Conclusion}
The increasing use of preprint servers makes it difficult for inexperienced researchers to distinguish between peer-reviewed quality research and uncontrolled preprints. Although most preprint servers support the option of linking preprints to published versions, the extent to which this option is used varies between research fields. For example, the experiments in this research identify a deficit in the field of computer science. The PreprintResolver has been developed to overcome this problem. The tool focuses on but is not limited to a target group of inexperienced researchers in the field of computer science. It uses a fuzzy matching algorithm to identify recent publications from arXiv preprints. A strength of the tool is that it uses four literature databases for resolving and the experiments show that all databases contributed to the final result. These literature databases are queried via their publicly available APIs, which guarantees the use of recent information. The tool requires minimal user interactions, making it more user-friendly. In addition, results are returned as BibTeX entries, allowing the direct integration with bibliographic tools. The PreprintResolver has been released as open source software. The experimental results show that the PreprintResolver was able to detect publications of 603 out of 1,000 arXiv preprints with missing publication information in the field of computer science. The resolving rate is stable across years with a sufficient number of samples in the dataset. A manual validation shows that the PreprintResolver identifies at least one plausible result for each of the 100 randomly sampled preprints. For nine samples, the tool identifies more than one result, and for three of these results, one of the results is incorrect or untraceable. This leads to the conclusion, that the PreprintResolver is suitable for individual, manually reviewed requests, but less suitable for bulk requests. Future work includes increasing the number of literature databases, which may increase the resolving ratio. In addition, the matching can be improved, for example, by training machine learning models for filtering of candidate matches. It is expected that such an approach will reduce the number of matches that are erroneously excluded by the strong filtering approach due to title or author changes prior to publication. Another idea is, to expand the supported preprint servers. Research building on this work will investigate the publication rate of arXiv preprints, the recall and precision of the PreprintResolver, and the role of each database on a larger dataset to improve the statistical validity. An extended analysis should also analyse the performance of different thresholds during the strong filtering. 

\section*{Acknowledgements}
This work is part of the BMBF-funded project ``Intelligente Unterstützung projekt- und problemorientierter Lehre und Integration in Studienabläufe'' (IPPOLIS) (Support code: 16DHBKI050).

The work of Louise Bloch was partially funded by a PhD grant from University of Applied Sciences and Arts Dortmund, Dortmund, Germany.

The authors thank arXiv for use of its open access interoperability. 

The authors thank the literature databases CrossRef / CrossCite, DBLP, SemanticScholar, and OpenAlex for the use of open access APIs and availability of data.

The authors thank SemanticScholar for providing an API key that allows requesting at a higher rate limit. 
%
%
 \bibliographystyle{splncs04}
 \bibliography{bibliography}
\end{document}